\documentclass[onecolumn,amsmath,amssymb,12pt,superscriptaddress,nofootinbib]{revtex4}
\pdfoutput=1

\usepackage[latin1]{inputenc}
\usepackage[english]{babel}
\usepackage{amssymb}
\usepackage{amsmath}
\usepackage{amsthm}
\usepackage[]{graphicx}
\usepackage[]{subfigure}
\usepackage{tensor}
\usepackage{color}
\usepackage{cancel}
\usepackage{setspace}
\usepackage{fancyhdr}
\usepackage[bookmarks,linktocpage, colorlinks=true, plainpages = false, citecolor = blue,  linkcolor=blue, urlcolor = blue, filecolor = blue]{hyperref}

\usepackage{amsmath,amsfonts,amssymb}
\usepackage{mathtools,physics}

\begin{document}
\allowdisplaybreaks
\begin{titlepage}
\begin{flushright}
Imperial/TP/2023/KSS/02
\end{flushright} 
\vspace{.2in}

\title{\Large{Higher-Order Gravity, Finite Action, and a \\ Safe Beginning for the Universe}
\vspace{.3in}}

\author{Jean-Luc Lehners}
\email{jlehners@aei.mpg.de}
\affiliation{Max--Planck--Institute for Gravitational Physics (Albert--Einstein--Institute), 14476 Potsdam, Germany}
\author{K.\,S. Stelle}
\email{k.stelle@imperial.ac.uk}
\affiliation{The Blackett Laboratory, Imperial College London, Prince Consort Road, London SW7 2AZ, U.K.}

\begin{abstract}
\vspace{.4in} \noindent 
General relativity allows for inhomogeneous and anisotropic universes with finite action. By contrast, in quadratic gravity such solutions obtain infinite action and are thus eliminated. What remains are homogeneous and isotropic solutions undergoing accelerated expansion, thereby automatically inducing an early inflationary phase. In this manner, semi-classical consistency may explain some of the basic, coarse-grained features of the early universe. This includes suitable initial conditions for the second law of thermodynamics, in the spirit of the Weyl curvature hypothesis. We note that quadratic gravity is a renormalisable theory and may admit an asymptotically safe regime at high energies, rendering the theory trustworthy to high energies. We also comment on theories containing curvature terms up to infinite derivative order, and on the contrast with no-boundary initial conditions.
\end{abstract}
\maketitle

\end{titlepage}

\tableofcontents


\section{Introduction}

The evolution of the universe presents a few very basic puzzles, which physicists have been aware of for many decades \cite{Dicke:1979jwq}. Gravity has a tendency to cause matter to clump, and for this reason it is puzzling that the universe is so homogeneous and isotropic on large scales. In fact, to still be homogeneous and isotropic to the extent observed today, the early universe must have been devoid of irregularities to very high precision. In terms of classical solutions to the equations of general relativity, obtaining this level of smoothness requires the use of extremely tuned initial conditions. Yet, in classical physics all initial conditions are {\it a priori} equally valid (and moreover, by Liouville's theorem, it does not even matter at what time they are analysed) -- so for this reason we may suspect that further understanding can only be obtained by developing a quantum theory of initial conditions. When an inflationary phase is added, the problem does not become fundamentally different \cite{Gibbons:2006pa}. This is because, although an inflationary phase dynamically smoothes the universe, it requires its own initial conditions, which once again are required to be rather special \cite{Ijjas:2013vea}. 

A complementary perspective on this issue was provided by Penrose, who framed his arguments in terms of the second law of thermodynamics \cite{Penrose:1979azm}. This law says that entropy/disorder grows over time. Hence, in the past the universe must have had a very low entropy, which is the same as saying that it must have been in a very special state. Penrose proposed that the Weyl curvature tensor (or more precisely its square) could be used as a quantitative measure of the entropy stored in the gravitational field. This led him to propose the {\it Weyl curvature hypothesis}, which is the premise that the Weyl curvature must have been vanishing (or very small, if one takes quantum fluctuations into account) at the initial stages of the universe.

Here we will review a set of ideas that show that, depending on the gravitational theory, very minimal semi-classical consistency requirements may already be sufficient to explain the special initial state of the universe, including in particular its homogeneity and isotropy \cite{Lehners:2019ibe}. More specifically, we will see that requiring saddle points of the gravitational path integral to have finite action can be enough to eliminate inhomogeneities and anisotropies, provided that gravity is described by a higher-order theory, the most interesting example being quadratic gravity. 

This idea has a precursor in the work of Barrow and Tipler, who first proposed the guiding principle that the action of the universe should be finite \cite{Barrow1988} (see also the update in \cite{Barrow:2019gzc}). The way they applied the idea was mainly rooted in classical physics, and for this reason they considered the action of the full $4-$dimensional universe, including its future evolution. This led to overly stringent requirements, which are difficult to reconcile with observations. Here, by contrast, we consider amplitudes from the early universe until present day field configurations, which seems a more appropriate framework to address cosmological observations. Moreover, this is also in the spirit of quantum gravity being seen foremost as an operational theory.

In the present overview, which may be regarded as a simplified, updated version of the ideas first presented in \cite{Lehners:2019ibe}, we will start by examining the consequences of finite action in ordinary general relativity in Section \ref{sec:gr}, finding that the principle is not very restrictive in that setting. Then, in Section \ref{sec:qg} we will examine quadratic gravity, where the finite action requirement is much more effective in selecting solutions, yet not so effective as to eliminate interesting solutions. We will also see how these ideas may fit with the concept of asymptotic safety. Subsequently, the analogous situation in gravitational theories with arbitrarily high orders of derivatives will be considered in Section \ref{sec:ao}, where we will also contrast the present setting with the no-boundary prescription. We will conclude in Section \ref{sec:discussion}.

\section{General relativity} \label{sec:gr}

It is instructive to first consider a few examples in ordinary general relativity. There, under rather general assumptions, the Hawking-Penrose theorem \cite{Hawking:1969sw} predicts the occurrence of an initial curvature singularity. It is then clear that the main danger of the action diverging stems from its behaviour close to this singularity. From the work of Belinsky, Khalatnikov and Lifshitz (BKL) \cite{Belinsky:1970ew} we know that the approach to the singularity involves brief periods during which the universe is (locally) well described by a Kasner metric. These Kasner epochs are interspersed with transition periods during which the Kasner anisotropies switch values, and moreover the directions of the Kasner axes can change. These changes occur chaotically and at ever increasing frequency in the approach to the singularity. The Kasner (Bianchi I) metric may be written as
\begin{align}
 \dd s^2_\mathrm{K}=-\dd t^2+a^2&\left[e^{\beta_++\sqrt{3}\beta_-}\dd x^2 +\,e^{\beta_+-\sqrt{3}\beta_-}\dd y^2+e^{-2\beta_+}\dd z^2\right]\,,\label{eq:B1}
\end{align}
with $a(t)$ being the average scale factor (defined such that the determinant of the metric is simply $\sqrt{-g} = a^3$) and $\beta_\pm(t)$ the anisotropies. The Einstein-Hilbert action then reduces to
\begin{align}
S_{EH} = \int d^4x \sqrt{-g} \frac{R}{2} = V_3 \int \dd t \,a \left( -3\dot{a}^2  + \frac{3}{4}a^2(\dot{\beta}^2_+ + \dot{\beta}^2_-) \right)\,,
\end{align}
where we have set $8\pi G=1$ and neglected surface terms (in the cases we consider these have the same convergence properties as the actions themselves). The spatial volume is denoted by $V_3$ above and we will assume that it is finite -- for example, the universe could have a toroidal topology\footnote{Obtaining a finite action requires a finite spatial volume, so one may say that an immediate prediction of semi-classical gravity is that the universe should be of finite spatial extent.}. The equations of motion and constraint resulting from this action are (with $H=\dot{a}/a$)
\begin{align} \label{Friedman}
0 &= \ddot\beta_\pm + 3H\dot\beta_\pm \,, \\
3H^2 &= \frac{3}{4}\dot{\beta}^2_+ + \frac{3}{4}\dot{\beta}^2_-  \,.
\end{align}
Thus the anisotropies behave as $\dot\beta_\pm \propto a^{-3}$ and the constraint then implies that 
\begin{align}
a(t)=a_0 t^{1/3}\,, \qquad \beta_\pm=b_\pm \ln (t)\,, \qquad \textrm{with } \, b_+^2+b_-^2=\frac{4}{9}\,.
\end{align}
Note that the anisotropies blow up in the approach to the singularity, $t \to 0,$ yet the on-shell action vanishes (as can be seen by plugging in the constraint) 
\begin{align}
S_{EH}^{on-shell} =0\,. \label{onshell}
\end{align}
This provides a simple example showing that in ordinary general relativity, singularities are common, and that they can easily have strong anisotropies yet finite action (and similar arguments can be made regarding inhomogeneities, see \cite{Lehners:2019ibe}). Semi-classically, such configurations are therefore expected to contribute significantly \cite{Feldbrugge:2017kzv}, but this is in conflict with observations of the early universe.

We can also include a coupling to matter, to see if this may prevent singularities from forming. We will consider the example of a perfect fluid with energy density $\rho$ and pressure $p=w\rho$, as many matter types of interest in cosmology may be modelled in that way. The action for the gravity plus perfect fluid system \cite{Schutz:1970my} can be written as
\begin{equation}
 S_{\textrm{fluid}}=\int\dd^4x\,\sqrt{-g}\,\left( \frac{R}{2}+\,p\right)\,.
\end{equation}
We will specialise to a  flat Friedmann-Lema\^itre-Robertson-Walker (FLRW) background metric here, given by
\begin{equation}
 \dd s^2_\mathrm{FLRW}=-\dd t^2+a(t)^2\dd\mathbf{x}^2\,.
\end{equation}
The continuity equation $\dot\rho + 3H(\rho + p)=0$ can be solved immediately to yield $\rho \propto a^{-3(1+w)},$ where we assume that the (constant) equation of state $w$ satisfies $w>-1.$ The constraint $3H^2=\rho$ then implies that 
\begin{equation}
 a \propto t^{\frac{2}{3(1+w)}}\,, \quad \rho \propto a^{-3(1+w)} \propto t^{-2}\,, \quad S_\textrm{fluid} \propto\int_0^{t_0} \dd t \, t^{-\frac{2w}{1+w}}\propto t_0^{\frac{1-w}{1+w}}-\lim_{t\to 0}t^{\frac{1-w}{1+w}}\,. \label{eq:fluid}
\end{equation}
The time integral in the action then converges as long as $-1<w\leq+1,$ yet for these values the curvature scalar $R=6\dot{H} + 12 H^2 \propto \frac{1}{t^2}$ clearly blows up. Hence once again we obtain singularities, yet finite action. 

A last example of cosmological relevance is that of a scalar field with potential $V(\phi),$ given by the action
\begin{equation}
S_{\phi}=\int\dd^4x\,\sqrt{-g}\left(\frac{R}{2}-\frac{1}{2}\partial_\mu\phi\partial^\mu\phi-V(\phi)\right)\,.\label{eq:SGRscalar}
\end{equation}
Variation with respect to the metric yields the Einstein field equations,
\begin{equation}
 R_{\mu\nu}-\frac{1}{2}Rg_{\mu\nu}=\partial_\mu\phi\partial_\nu\phi+g_{\mu\nu}\left(-\frac{1}{2}\partial_\alpha\phi\partial^\alpha\phi-V(\phi)\right)\,,\label{eq:EEs}
\end{equation}
whose trace is given by
\begin{equation}
R=-T=\partial_\mu\phi\partial^\mu\phi+4V(\phi)\,.
\end{equation}
Substituting the trace, we find the on-shell action 
\begin{equation}
 S_{\phi,\textrm{on-shell}}=\int\dd^4x\,\sqrt{-g}\,V(\phi)\,. \label{eq:scalaronshell}
\end{equation}
The action will certainly converge when both the potential and the $4$-volume $\int \dd^3x\,\dd t\,\sqrt{-g}$ remain bounded \cite{Jonas:2021xkx}. However, the action can be finite even if the potential blows up at the big bang, as long as it does not blow up too fast. An explicit example is given by the class of exponential potentials  $V(\phi)=V_0 \, e^{\sqrt{2/s}\, \phi},$ which yield the exact FLRW solutions $a(t)\propto t^s$ and $\phi(t)=-\sqrt{2s}\ln(t\sqrt{V_0/[s(3s-1)]}),$ where we must impose $s>1/3.$ All these solutions have curvature singularities at $t=0,$ yet the on-shell action converges since $V(\phi) \propto t^{-2}$ and $a^3 \propto t^{3s}$ with $3s>1.$ 

All these examples show that in ordinary general relativity, solutions containing singularities yet yielding finite action are ubiquitous.

\section{Quadratic gravity and Asymptotic Safety} \label{sec:qg}

We can now extend our analysis to quadratic gravity, which contains terms up to four derivatives. When coupled to the Einstein-Hilbert theory, the action becomes \cite{Stelle:1977ry}
\begin{align}
S_{\textrm{quad}} & = \int d^4x \sqrt{-g} \left[ \frac{1}{\kappa^2} R -\frac{1}{2\sigma} C_{\mu\nu\rho\sigma}C^{\mu\nu\rho\sigma} + \frac{\omega}{3\sigma} R^2  \right]\,, \label{QuadGravity}
\end{align}
where $\kappa, \sigma, \omega$ are coupling constants which we will discuss in more detail below. $C_{\mu\nu\rho\sigma}$ denotes the Weyl tensor.

We would like to see if big bang singularities also occur with finite action in this theory. For this, we will first revert to the BKL/mixmaster setting described above. We will refine our analysis slightly, by considering the Bianchi IX metric instead of the Kasner metric. This is because, as shown by BKL \cite{Belinsky:1970ew}, the Bianchi IX metric is locally ``generic'' in the sense that it involves four free functions. It thus describes the dynamics near a cosmological singularity rather accurately, also in the presence of ordinary matter. The metric may be seen as a curved generalisation of the Kasner metric \eqref{eq:B1}, and is given by
\begin{align}
 \dd s^2_\mathrm{IX}=-N^2 \dd t^2+a^2&\left[e^{(\beta_++\sqrt{3}\beta_-)}\sigma_1^2 +\,e^{(\beta_+-\sqrt{3}\beta_-)}\sigma_2^2+e^{-2\beta_+}\sigma_3^2\right]\,,\label{eq:B9}
\end{align}
with the one-forms $\sigma_1 = \sin\psi\, d\theta - \cos \psi \sin \theta\, d\varphi$, $\sigma_2 = \cos \psi\, d\theta + \sin \psi \sin \theta\, d \varphi$, and $\sigma_3 = -d\psi + \cos\theta\, d\varphi$ and coordinate ranges $0 \leq \psi \leq 4 \pi$, $0 \leq \theta \leq \pi$, and $0 \leq \phi \leq 2 \pi.$  When $\beta_- = \beta_+ = 0$ one recovers the FLRW metric with closed spatial sections. Above, we have also included the lapse function $N,$ which we will take to be a constant.

The Einstein-Hilbert action then reduces to
\begin{align}
S_{EH} = \int d^4x \sqrt{-g} \frac{R}{2} = 2\pi^2 \int N \dd t a \left( -3\frac{\dot{a}^2}{N^2}  + \frac{3a^2}{4N^2}(\dot{\beta}^2_+ + \dot{\beta}^2_-) -  U(\beta_+, \beta_-)\right)\,,
\end{align}
where the anisotropy parameters obtain an effective potential of the form
\begin{align} \label{anisotropypotential}
U(\beta_+, \beta_-)  = - 2 \left( e^{ 2 \beta_+ } + e^{-\beta_+ - \sqrt{3}\beta_-} + e^{-\beta_+ + \sqrt{3}\beta_-} \right) + \left( e^{ -4 \beta_+ } + e^{2\beta_+ - 2\sqrt{3}\beta_-} + e^{2\beta_+ + 2\sqrt{3}\beta_-} \right)\,.
\end{align}
Meanwhile, the $(\hbox{Weyl})^2$ part of the action reads (up to total derivative terms)
\begin{align}
& - \int \dd^4x \sqrt{-g}\, C_{\mu\nu\rho\sigma}C^{\mu\nu\rho\sigma} \nonumber \\ = & \, 2\pi^2 \int N \dd t   \Big{\{} 3 \frac{a^3}{N^4} \left[\left( \frac{\ddot{a}}{a}+H^2 \right) (\dot\beta_{-}^2 + \dot\beta_{+}^2) -\ddot\beta_-^2 - \ddot\beta_+^2 -(\dot\beta_{-}^2 + \dot\beta_{+}^2)^2\right]  \nonumber \\ &  \, + 4\frac{a}{N^2} \left[ -(\ddot\beta_- + 3H \dot\beta_-) U_{,\beta_-} -(\ddot\beta_+ + 3H \dot\beta_+) U_{,\beta_+} -\left(2\frac{\ddot{a}}{a} + \dot\beta_-^2 + \dot\beta_+^2\right)U\right] \nonumber \\ &  \, + \frac{64}{3a} \left( - e^{-8\beta_+} + e^{-5\beta_+ -\sqrt{3}\beta_-} + e^{-5\beta_+ + \sqrt{3}\beta_-} - e^{-2 \beta_+} + e^{\beta_+ -3\sqrt{3} \beta_-} + e^{\beta_+ + 3 \sqrt{3} \beta_-}  \right.\nonumber \\ &  \left. \qquad \quad- e^{\beta_+ - \sqrt{3}\beta_-} - e^{\beta_+ + \sqrt{3}\beta_-} - e^{4\beta_+ -4\sqrt{3} \beta_-} - e^{4\beta_+ + 4 \sqrt{3}\beta_-}+ e^{4\beta_+ -2\sqrt{3} \beta_-} + e^{4\beta_+ +2\sqrt{3}\beta_-}\right) \Big\} \,,\label{actionweylsquared}
\end{align}
while the Ricci scalar squared contribution is 
\begin{align}
  \int d^4x \sqrt{-g} R^2
=   \, 2\pi^2 \int N \dd t  a^3 \left[ 6 \frac{\ddot{a}}{N^2a} + 6\frac{\dot{a}^2}{N^2a^2}  + \frac{3}{2N^2}(\dot{\beta}^2_+ + \dot{\beta}^2_-) -  \frac{2}{a^2}U(\beta_+, \beta_-)\right]^2\,. \label{actionRsquared}
\end{align}

As we approach the singularity, $a \to 0,$ the ``dangerous'' terms in the $(\hbox{Weyl})^2$ and Ricci squared actions are seen to be of the form $\int N \dd t \, \frac{U^2}{a}$ and $\int N \dd t \, \frac{u(\beta_\pm)}{a}$ for some function $u(\beta_\pm)$ that can be read off from the $(\hbox{Weyl})^2$ action. We can use the constraint to eliminate these terms from the on-shell action. The constraint is obtained by calculating a derivative with respect to the lapse $\frac{\delta}{\delta N},$ so that it involves the terms shown in the action, but with different relative coefficients depending on whether their lapse dependence is $\frac{1}{N^3}, \frac{1}{N}$ or proportional to $N.$ This implies that after we eliminate the dangerous terms above, we will be left with terms proportional to  $\frac{1}{N^3}, \frac{1}{N},$ albeit with different numerical coefficients. The on-shell action will thus include integrands of the form $ a^2 \ddot{a} \dot\beta^2, \, a^3 \ddot\beta^2$ and similar. The exponential nature of the potential implies that the anisotropy parameters $\beta_\pm$ will continue to evolve logarithmically in time, or $\dot\beta_\pm \propto 1/t.$ Then if we again write an ansatz $a \propto t^s,$ we find that the on-shell action scales as
\begin{align}
S_{on-shell} \propto \int_0^{t_0} \dd t \, t^{3s-4} \propto t_0^{3s-3}-\lim_{t\to 0}t^{3s-3}\,.
\end{align}
For convergence we require $s>1$ -- in other words we require an \emph{accelerating} solution. This is significantly different than the case of ordinary general relativity, for two reasons: first, we may recall that the BKL analysis showed that in the approach to a big bang, the scale factor behaves approximately as $a \propto t^{1/3}$ (or, to put it differently, the spatial volume shrinks as $t$). This means that the standard BKL mixmaster behaviour is eliminated. And secondly, when the universe is accelerating, anisotropies are damped away fast, even if they are present initially. In this way, quadratic gravity automatically selects inflationary initial conditions and dilutes both anisotropies and inhomogeneities.

We can consolidate this finding by looking at the coupling to matter. Here we will restrict our analysis to a perfect fluid, which we will once again take to have energy density $\rho$ and pressure $p = w\rho.$ The continuity equation remains applicable in higher-order gravity theories \cite{DeFelice:2010aj}, so that we have 
\begin{align}
\dot\rho + 3H(\rho + p)=0 \rightarrow \rho = \frac{\rho_0}{a^{3(1+w)}}\,,
\end{align}
where we assumed a FLRW metric. For the gravitational part, we will take an action of the form
\begin{align}
S_n = \frac{1}{2}\int \dd^4x \sqrt{-g} \left[ f(R) + p\right]\,, \label{fR}
\end{align}
where we will take $f(R)$ to be a polynomial with highest-order term $R^n.$ We also write $F=f_{,R}.$ Then the constraint reads \cite{DeFelice:2010aj}
\begin{align}
3FH^2 = \frac{1}{2}(FR-f) -3H\dot{F} + \rho\,. \label{highconstraint}
\end{align}
With the ansatz that $a \propto t^s,$ we have that
\begin{align}
H=\frac{s}{t}\,, \quad R = 6\dot{H} + 12 H^2 = \frac{12s^2-6s}{t^2}\,.
\end{align}
Hence in \eqref{highconstraint} the leading gravity terms scale as $t^{-2n}$ as $t \to 0,$ while the energy density scales as $\rho \propto t^{-3s(1+w)}.$ For a solution to exist, we must take
$s=\frac{2n}{3(1+w)}$
and hence the action scales as
\begin{align}
S_n  \sim \frac{V_3}{2}\int_0^{t_0} \dd t \, a^3 R^n \propto t_0^{\frac{2n}{1+w}-2n+1} - \lim_{t\to 0} t^{\frac{2n}{1+w}-2n+1}\,.
\end{align}
For convergence we therefore require 
\begin{align}
w<\frac{1}{2n-1}\, \qquad s> \frac{2n-1}{3}\,.
\end{align}
For $n=1$ we recover the general relativity result $w<1,$ while for quadratic gravity ($n=2$) we find $w<\frac{1}{3}$ or $a(t) \propto t^s$ with $s>1,$ confirming  that quadratic gravity selects accelerated expansion in the approach to the big bang.

Let us now offer some comments on the significance and trustworthiness of this observation. The important feature is that quadratic gravity is a renormalisable theory, due to the $1/k^4$ momentum dependence of the propagator at high energies \cite{Stelle:1976gc}. This implies that the couplings in the action \eqref{QuadGravity} will run with energy scale, but no new terms of higher order in derivatives will be generated by quantum corrections\footnote{In addition to massless gravitons, the theory contains new degrees of freedom: negative energy gravitons of mass squared $\sigma/\kappa^2 \sim \mu^2/\ln\mu$ and scalar excitations of mass squared $\frac{\sigma}{2\kappa^2 \omega} \sim \mu^2/\ln\mu$ \cite{Stelle:1977ry}. The impact of the graviton ghost remains a matter of debate, to which we have no new ideas to add here, except to point out that some protection from the ghost may arise due to its mass becoming large as the energy scale $\mu$ grows.}. In fact, if we introduce an energy scale $\mu$ and consider the addition of a cosmological constant $\Lambda,$ then we may define dimensionless couplings $\frac{1}{\kappa^2} = \frac{\mu^2}{g_N}$ and $\Lambda = \lambda \mu^4.$ The other two couplings $\sigma, \omega$ are dimensionless by definition. The running of these couplings has been calculated (though so far only in a Euclidean setting), with the result that \cite{Fradkin:1981iu,Avramidi:1985ki,Codello:2006in,Niedermaier:2009zz,Niedermaier:2010zz,Ohta:2013uca}
\begin{align}
\mu \frac{d}{d\mu} g_N & = f_g(g_N,\lambda,\sigma,\omega)\,, \qquad \qquad \mu \frac{d}{d\mu} \lambda  =  f_\lambda(g_N,\lambda,\sigma,\omega)\,, \nonumber \\
\mu \frac{d}{d\mu} \sigma & = - \frac{133}{160\pi^2}\sigma^2\,,  \qquad \quad \,\, \mu \frac{d}{d\mu} \omega  = - \frac{25+1098 \omega + 200 \omega^2}{960\pi^2} \sigma\,.
\end{align}
In fact, there is evidence for a fixed point at high energies, at $\sigma^\star=0,\omega^\star\approx -0.0228$ and at finite positive (but scheme dependent) values $g_N^\star, \lambda^\star,$ with  $\sigma \sim 1/\ln\mu$ decaying inversely to the logarithm of the energy scale \cite{Codello:2006in,Niedermaier:2009zz,Niedermaier:2010zz,Ohta:2013uca} (although other works find a non-trivial fixed point at non-zero values of  $\sigma, \omega$ \cite{Benedetti:2009rx}). This would realise Weinberg's idea of asymptotic safety \cite{Weinberg:1980gg} and would imply that we may be able to take this theory seriously up to arbitrarily high energies. In that case, it is justified to take the action integrals all the way to vanishing scale factor $a=0,$ as we have done.

\section{All-orders gravity and relation to the no-boundary proposal} \label{sec:ao}

If instead of quadratic gravity one starts with ordinary general relativity and adds in quantum corrections, then one finds that loop diagrams contribute terms of higher order in derivatives. In fact, since general relativity is not renormalisable  \cite{tHooft:1974toh,Christensen:1979iy,Goroff:1985th}, terms up to arbitrarily high order are generated in this way. Their couplings will run with energy, and the question arises as to whether a non-trivial, asymptotically safe, fixed point may exist \cite{Reuter:1996cp,Percacci:2017fkn,Eichhorn:2018yfc}. If so, then one would expect the theory at the fixed point to exhibit scale invariance, which may imply that the properties of such a theory might be similar to those of quadratic gravity (or pure $(\hbox{Weyl})^2$ gravity). In that case, one would recover the results obtained in the previous section. But if no such fixed point turns out to exist, or if one truncates the theory at some finite order, one may still wander what happens to the arguments discussed above.

For this purpose, let us reconsider the $f(R)$ theory coupled to a perfect fluid, discussed from Eq.~\eqref{fR} onwards, and containing terms up to order $R^n$. There we found that, for a fluid with equation of state $w,$ we obtain a convergent action only if
\begin{align}
w<\frac{1}{2n-1}\, \qquad s> \frac{2n-1}{3}
\end{align}
where the scale factor evolves as $a(t) \propto t^s$ as $t \to 0.$ Clearly, the higher the power of $R,$ the more stringent the requirements. At finite $n$ one requires the scale factor to undergo a phase of super-acceleration, but here we should keep in mind that truncating the theory at finite $n > 2$ still leads to a non-renormalisable theory. Moreover, such truncated theories are all plagued by ghost states. This motivates us to look at the limit where we keep terms up to infinite order. However, it is immediately apparent from the inequalities above that no non-singular solution may be found in such a case. In other words, if one keeps the full series of Riemann terms, then all Lorentzian big bang spacetimes are eliminated semi-classically \cite{Jonas:2021xkx}.

The focus on Lorentzian spacetimes is important here, as there exists a way of circumventing this no-go result, and which consists in enlarging the class of spacetimes that one is willing to consider. This alternative way of fixing initial conditions is the no-boundary proposal \cite{Hartle:1983ai} (for a review see \cite{Lehners:2023yrj}). The idea of the no-boundary proposal is to consider spacetimes that are rounded off (like the surface of a ball, but in $4$ dimensions) near the putative big bang. In other words, the metric should be both compact and regular everywhere. If such a metric exists, then by construction it will lead to finite action, and stands a chance of being relevant semi-classically. 

The simplest metric we can consider is again of FLRW type, but with closed spatial sections,
\begin{equation}
 \dd s^2=-\dd t^2+a(t)^2\dd\Omega_3^2\,.
\end{equation}
where $\dd \Omega_3^2$ is the metric on the $3-$sphere. However, in contrast to the situation in classical physics, we will allow the scale factor $a(t)$ to take complex values here (while the coordinate $t$ remains real valued). We will again consider the toy model theory in Eq.~\eqref{fR},
with a polynomial function $f(R)=\sum_{m=0}^{n} c_m R^m.$ Then the constraint (time-time Einstein equation) becomes \cite{DeFelice:2010aj}
\begin{align}
\rho & = \left( 3H^2+\frac{3}{a^2}\right) F + \frac{1}{2}\left( f-RF\right)  +3H\dot{F}&\\
& = \sum_{m=0}^{n} c_m \left[\left( 3H^2+\frac{3}{a^2}\right) mR^{m-1} +\frac{1}{2}(1-m)R^m + 3m(m-1)HR^{m-2}\dot{R} \right] \label{eqcon}
\end{align}
The idea of the no-boundary proposal is that $a(t)$ in fact starts out pure imaginary near $t=0,$ and this point should now be viewed like a point on the surface of a ball (an analogy would be the South Pole on Earth). The ansatz for the scale factor reads 
\begin{align}
a(t) &= it + \frac{a_3}{6}t^3 + \frac{a_5}{120}t^5 + \cdots \,,
\end{align}
where the linear term $it$ ensures the regularity of the solution at the origin. Here we have used the fact that the expansion of $a$ contains only odd powers, a result which is straightforward to derive and for which we refer to \cite{Lehners:2023yrj}. Then the Ricci scalar becomes
\begin{align}
R & = 6\left(\frac{\ddot{a}}{a} + \frac{\dot{a}^2}{a^2} + \frac{1}{a^2} \right) = 12i a_3 + \frac{3}{2}\left( a_3^2 + i a_5\right) t^2 + \cdots\,. \label{eqR}
\end{align}
The interesting observation is that the Ricci scalar, and by extension also the right hand side of \eqref{eqcon}, starts at order ${\cal O}(t^0),$ and in particular does not contain negative powers of $t,$ which would have led to a divergence. The action remains manifestly finite, as the curvature does not blow up, and its time integral is over a finite range. Allowing a Euclidean ``beginning'' thus allows for finite action solutions, even if the series is truncated at a finite order $n.$ 

Full solutions are then obtained by solving the constraint order by order in time. At zeroth order, for the example where $f=R^n$, one obtains
\begin{align}
\rho(t=0) = 3^n (4ia_3)^{n-2} \left[a_3^2(n^2+3n-8)+in(n-1)a_5\right] \,.
\end{align}
Hence the coefficient $a_5$ is given in terms of the coefficients $c_m$ and $a_3,$ while higher order equations progressively fix higher coefficients in the expansion of $a(t)$ in terms of the lower ones. The coefficient $a_3$ remains free, and may be understood as a parameter labelling different solutions, with different expansion rates and in fact also different weightings -- since solutions will be complex, so will the action, and different solutions are then weighted in proportion to $e^{-Im(S)/\hbar}.$ In fact it is precisely in this manner that the no-boundary proposal provides a quantum (probabilistic) theory of initial conditions. The weighting usually favours small expansion rates, but there can be situations where a minimum rate is required in order for the solution to be stable and in such cases agreement with experiment seems possible~\cite{Lehners:2022mbd,Lehners:2022xds,Hertog:2023vot,Lehners:2023pcn}.

The toy model calculation above can be extended to theories with different contractions of the Riemann tensor, and also specific gravitational theories obtained in low energy string theory \cite{Jonas:2020pos}. One important difference with the finite action Lorentzian solutions discussed earlier is that in the no-boundary case the curvature remains finite and no arbitrarily high energy scale is reached. In that sense one would never reach an asymptotically safe fixed point in the description of the creation of the universe, but unbounded energy scales could of course still occur in future (black hole) singularities \cite{Borissova:2020knn}.

\section{Discussion} \label{sec:discussion}

We have seen that semi-classical consistency, or more precisely the requirement that relevant saddle points of the gravitational path integral have finite action, acts as a sieve and enacts a strong \emph{selection principle} on spacetimes. This principle is especially beneficial in quadratic gravity, where it filters out initially non-accelerating spacetimes. The remaining, accelerating, solutions thus automatically initiate a period of cosmic inflation, in which anisotropies and inhomogeneities are quickly ironed out. In this sense Penrose's Weyl curvature hypothesis is implemented, providing an appropriate starting point for the second law of thermodynamics. What is more, quadratic gravity is renormalisable and may admit an ultraviolet fixed point under the renormalisation group flow, implying that one may be able to trust the theory up to arbitrarily high energy scales.

Even though the scenario envisaged here works particularly well in quadratic gravity, it may have wider applicability. On the one hand, this is because if ordinary general relativity also happens to be asymptotically safe, then the effective theory at/near the fixed point may be closely related to quadratic gravity, due to its scale invariance. On the other hand, asymptotically safe gravity may not be the ultimate theory of quantum gravity, but may represent an intermediate energy regime with, for instance, string theory providing the fundamental theory of quantum gravity in the deep ultraviolet \cite{deAlwis:2019aud}. In such a case, our results may describe the effective genesis of spacetime from different, potentially even non-geometric, degrees of freedom.

We have also described the contrast with no-boundary initial conditions, which may be imposed in a wide variety of gravitational theories. By construction, no-boundary solutions satisfy the principle of finite action, and provide a different, non-Lorentzian implementation thereof. It may be interesting to point out that our scenario of a safe beginning may be seen as selecting either theories or solutions, or both (a point also made in \cite{Borissova:2020knn}), while the no-boundary proposal is less theory-specific and provides probabilities for different beginnings, as long as a source of positive vacuum energy is present. Distinguishing between the two proposals on an observational level is an interesting challenge for future research.

There exist a number of avenues for further exploration: it would be good to clarify the relation between energy scale and spacetime curvature, in generic spaetimes. This would allow for a more precise characterisation of the induced inflationary phase, and would clarify whether this phase lasts long enough by itself, or whether the addition of an inflaton field is required. Also, the calculations of non-trivial fixed points have so far mainly been performed in a Euclidean setting, and it would be important to study whether this is justified, and whether the results can unambiguously be Wick rotated to Lorentzian spacetimes. Furthermore, we have provided a number of arguments using specific, symmetry-reduced forms of the metric. It seems possible however to generalise some of the results to arbitrary metrics, perhaps by using a canonical $(3+1)$ decomposition of the metric \cite{Arnowitt:1962hi}. We leave these questions for future work.

In concluding, let us point out a historical curiosity: The papers by Dicke and Peebles highlighting the puzzles of the hot big bang model \cite{Dicke:1979jwq}, Penrose's discussion of the second law of thermodynamics and his Weyl curvature hypothesis \cite{Penrose:1979azm}, Weinberg's idea of asymptotic safety in gravity \cite{Weinberg:1980gg}, and finally Hawking's ideas regarding the path integral quantisation of gravity using Euclidean metrics \cite{Hawking}, all appeared side by side in the same Einstein Centenary Survey in 1979. Clearly, they were destined to eventually be combined together.

\acknowledgments

We would like to thank Caroline Jonas, Jerome Quintin, Rahim Leung and the late John Barrow for helpful discussions and correspondence. JLL gratefully acknowledges the support of the European Research Council in the form of the ERC Consolidator Grant CoG 772295 ``Qosmology''. The work of KSS was supported in part by the STFC under Consolidated Grants ST/T000791/1 and ST/X000575/1. KSS would like to thank the Albert Einstein Institute and the Perimeter Institute for hospitality during the course of the work.

\bibliographystyle{utphys}
\bibliography{SafeBeginning2}

\end{document}